\documentclass[fleqn,12pt]{wlscirep}
\usepackage{euscript,amsmath,amssymb,amsfonts,graphicx,bm,cite,appendix,cite}
  \usepackage{paralist}
  \usepackage{epstopdf}
  \usepackage{graphics} 
  \usepackage{enumitem}
  \usepackage{boxedminipage}
 \usepackage[colorlinks=true]{hyperref}
 \hypersetup{urlcolor=blue, citecolor=red}
 
 \usepackage{chngcntr}
\counterwithout{figure}{section}
\counterwithout{equation}{section}

\renewcommand{\d}{{\rm d}}
\renewcommand{\P}{{\rm P}}

\setlist[itemize]{leftmargin=*}

\title{Optimal models of decision-making in dynamic environments}

\author[1]{Zachary P. Kilpatrick}
\author[2]{William R. Holmes}
\author[1]{Tahra L. Eissa}
\author[3,4]{Kre\v{s}imir Josi\'{c}}

\affil[1]{Department of Applied Mathematics, University of Colorado, Boulder, Colorado, USA} 
\affil[2]{Department of Physics and Astronomy, Department of Mathematics, Quantitative Systems Biology Center, Vanderbilt University, Nashville, Tennessee, USA}
\affil[3]{Department of Mathematics, Department of Biology and Biochemistry, University of Houston, Houston, Texas, USA}
\affil[4]{Department of BioSciences, Rice University, Houston, Texas, USA}
\affil[*]{zpkilpat@colorado.edu, josic@math.uh.edu}


\begin{abstract}

Nature is  in constant flux, so animals must account for changes in their environment when making decisions. How animals learn the timescale of such changes and adapt their decision strategies accordingly is not well understood. Recent psychophysical experiments have shown humans and other animals can achieve near-optimal performance  at two alternative forced choice (2AFC) tasks in dynamically changing environments. Characterization of performance requires the derivation and analysis of computational models of optimal decision-making policies on such tasks. We review recent theoretical work in this area, and discuss how models compare with subjects' behavior in tasks where the correct choice or evidence quality changes in dynamic, but predictable, ways.

\end{abstract}

\begin{document}

\flushbottom
\maketitle  

\thispagestyle{empty}

\section*{Introduction}

To translate stimuli into decisions, animals interpret sequences of observations based on their prior experiences~\cite{gold07}. 
However, the world is fluid: The context in which a decision is made, the quality of the evidence, and even the best choice can change before  a judgment is formed, or an action taken. A source of water can dry up,  or a nesting site can become compromised. But even when not fully predictable, changes often have statistical structure: Some changes are rare, others are frequent, and some are more likely to occur at specific times. How have animals adapted their decision strategies to a world that is structured, but in flux?

Classic computational, behavioral, and neurophysiological studies of decision-making mostly involved tasks with fixed or statistically stable evidence~\cite{britten92,bogacz06,gold07}. To characterize the neural computations underlying decision strategies in changing environments, we must understand the dynamics of evidence accumulation~\cite{gao17}.  
This requires novel theoretical approaches. While normative models are a touchstone for theoretical studies~\cite{glaze15,piet18}, even for simple dynamic tasks the computations required to optimally translate evidence into decisions can become prohibitive~\cite{adams07}.
Nonetheless, quantifying how behavior differs from normative predictions helps elucidate the assumptions animals use to make decisions~\cite{radillo17,glaze18}. 

We review normative models and compare them with experimental data from two alternative forced choice (2AFC) tasks in dynamic environments.  Our focus is on tasks where subjects passively observe streams of evidence, and the evidence quality or correct choice can vary within or across trials.  Humans and animals adapt their decision strategies to account for such volatile environments, often resulting in performance that is nearly optimal on average.  However, neither the computations they use to do so nor their neural implementations are well understood.


\section*{Optimal evidence accumulation in changing environments}

\begin{figure}
\begin{center} \includegraphics[width=15cm]{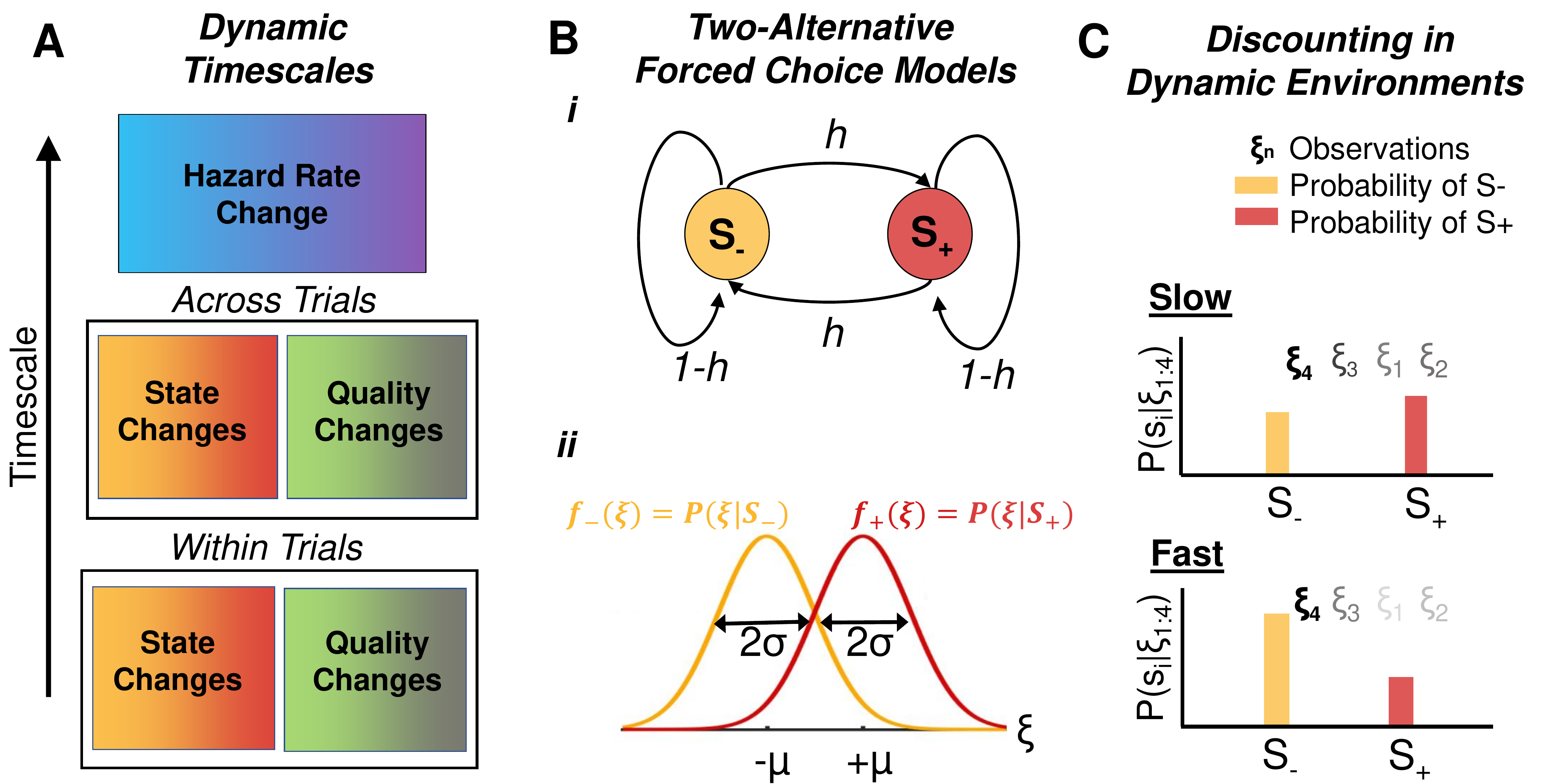} \end{center}
\vspace{-2mm}
\caption{{\bf Two alternative forced choice (2AFC) tasks in dynamic environments.}  {\bf (A)} Possible timescales of environmental dynamics: 1) The state ($S_+$ or $S_-$), or the quality of the evidence (e.g., coherence of random dot motion stimulus) may switch within a trial~\cite{glaze15,krishnamurthy17,piet18}, or 2) across trials~\cite{goldfarb12,purcell16,malhotra17}; 3) The hazard rate, $h$, can change across blocks of trials~\cite{piet18,glaze18}. {\bf (B)} In a dynamic 2AFC task, a two-state Markov chain with hazard rate $h$ determines the state. {\bf (Bi)} The current state (correct hypothesis) is either $S_+$ (red) or $S_-$ (yellow).  {\bf (Bii)} Conditional densities of the observations, $f_{\pm}(\xi) = P(\xi | S_{\pm}),$ shown as Gaussians with means $\pm\mu$ and standard deviation $\sigma$. {\bf (C)} Evidence discounting is shaped by the environmental timescale: (Top) In slow environments, posterior probabilities over the states, $P(S_{\pm}| \xi_{1:4} ),$ are more strongly influenced by past observations, $\xi_{1:3}$,  (darker shades of the observations, $\xi_i$, indicate higher weight) and thus points to $S_+$. (Bottom) If changes are fast, beliefs depend more strongly on the current observation, $\xi_4$, which outweighs older evidence and points to $S_-$.}   
\label{fig1:schematic}
\end{figure}

Normative models of decision-making typically assume subjects are Bayesian agents~\cite{Geisler03,Knill04} that probabilistically
compute their belief of the state of the world by combining fresh evidence with previous knowledge. Beyond normative models, notions of optimality require a defined objective. For instance, an observer may need to report the location of a sound~\cite{kim17}, or the direction of a moving cloud of dots~\cite{glaze15}, and is  rewarded if the report is correct. 
Combined with a  framework to translate probabilities or beliefs into actions, normative models provide a rational way to maximize the net rewards dictated by the environment and task. Thus an optimal model combines normative computations  with a policy that translates a belief into the optimal action. 

How are normative models and optimal policies in dynamic environments characterized? Older observations have less relevance in rapidly changing environments than in slowly changing ones. Ideal observers account for environmental changes by adjusting the rate at which they discount prior information when making inferences and decisions~\cite{velizcuba16}.  In Box \textcolor{blue}{1} we show how, in a normative model, past evidence is nonlinearly discounted at a rate  dependent on environmental volatility~\cite{glaze15,velizcuba16}. 
When this volatility~\cite{radillo17} or the underlying evidence quality~\cite{drugowitsch12,malhotra17} are unknown, they must also be inferred. 

In 2AFC tasks, subjects accumulate evidence until they decide on one of two choices either freely or when interrogated.
In these tasks, fluctuations can act on different timescales (Fig.~\ref{fig1:schematic}a): 
 1) on each trial (Fig.~\ref{fig1:schematic}b,c)~\cite{glaze15,piet18}, 2) unpredictably within only some trials~\cite{holmes2016new,holmes2018bayesian}, 3) between trials in a sequence~\cite{goldfarb12,kim17}, or 4) gradually across long blocks of trials~\cite{Yu08}. We review findings in the first three cases and compare them to predictions of normative model.

\begin{figure}[t]
\begin{boxedminipage}{\textwidth}
\noindent
{\bf Box 1 -- Normative evidence accumulation in dynamic environments.} \\
{\em Discrete time.} At times $t_{1:n}$ an observer receives a sequence of noisy observations, $\xi_{1:n},$ of the state $S_{1:n}$, governed by a two-state Markov process (Fig.~\ref{fig1:schematic}b). Observation likelihoods, $f_{\pm} (\xi) = \P (\xi | S_{\pm})$, determine the belief (log-likelihood ratio: LLR), $y_n = \log \frac{\P(S_n = S_+ | \xi_{1:n})}{\P(S_n = S_- | \xi_{1:n})},$ after observation $n$. If the observations are conditionally independent, the LLR can be updated recursively~\cite{glaze15,velizcuba16}:
\begin{align}
y_n = \underbrace{\log \frac{f_+(\xi_n)}{f_-(\xi_n)}}_{\text{current evidence}} + \underbrace{\log \frac{(1-h) \exp ( y_{n-1}) + h}{h \exp ( y_{n-1}) + (1-h)}}_{\text{discounted prior belief}}, \label{LLR1}
\end{align}
where $h$ is the hazard rate (probability the state switches between times $t_{n-1}$ and $t_n$). The belief  prior to the observation at time $t_n$, $y_{n-1}$, is discounted according to the environment's volatility $h$. When $h = 0$, Eq.~(\ref{LLR1}) reduces to the classic drift-diffusion model (DDM), and evidence is accumulated perfectly over time. When $h= 1/2$, only the latest observation, $\xi_n,$ is informative.  For $0<h<1/2$, prior beliefs are discounted, so past evidence contributes less to the current belief, $y_n$, corresponding to leaky integration. When $1/2<h<1$, the environment alternates. \\
\vspace{-4mm}

{\em Continuous time.} When  $t_n - t_{n-1} = \Delta t \ll 1$, and the hazard rate is defined $\Delta t \cdot h$, LLR evolution can be approximated by the  stochastic differential equation~\cite{glaze15,velizcuba16}:
\begin{align}
\d y = \underbrace{g(t) \d t}_{\text{drift}} + \underbrace{\d W_t}_{\text{noise}} -  \underbrace{2 h \sinh (y) \d t}_{\text{nonlinear filter}}, \label{SDE}
\end{align}
where $g(t)$ jumps between  $+ g$ and $-g$ at a rate $h$, $W_t$ is a zero mean Wiener process with variance $\rho^2$, and the nonlinear filter $-2 h \sinh(y)$ optimally discounts prior evidence. In contrast to the classic continuum DDM, the belief, $y(t),$ does not increase indefinitely, but  saturates due to evidence-discounting.  
\end{boxedminipage}
\end{figure}

\section*{Within trial changes promote leaky evidence accumulation}

\begin{figure}[t]
\begin{center} \includegraphics[width=15cm]{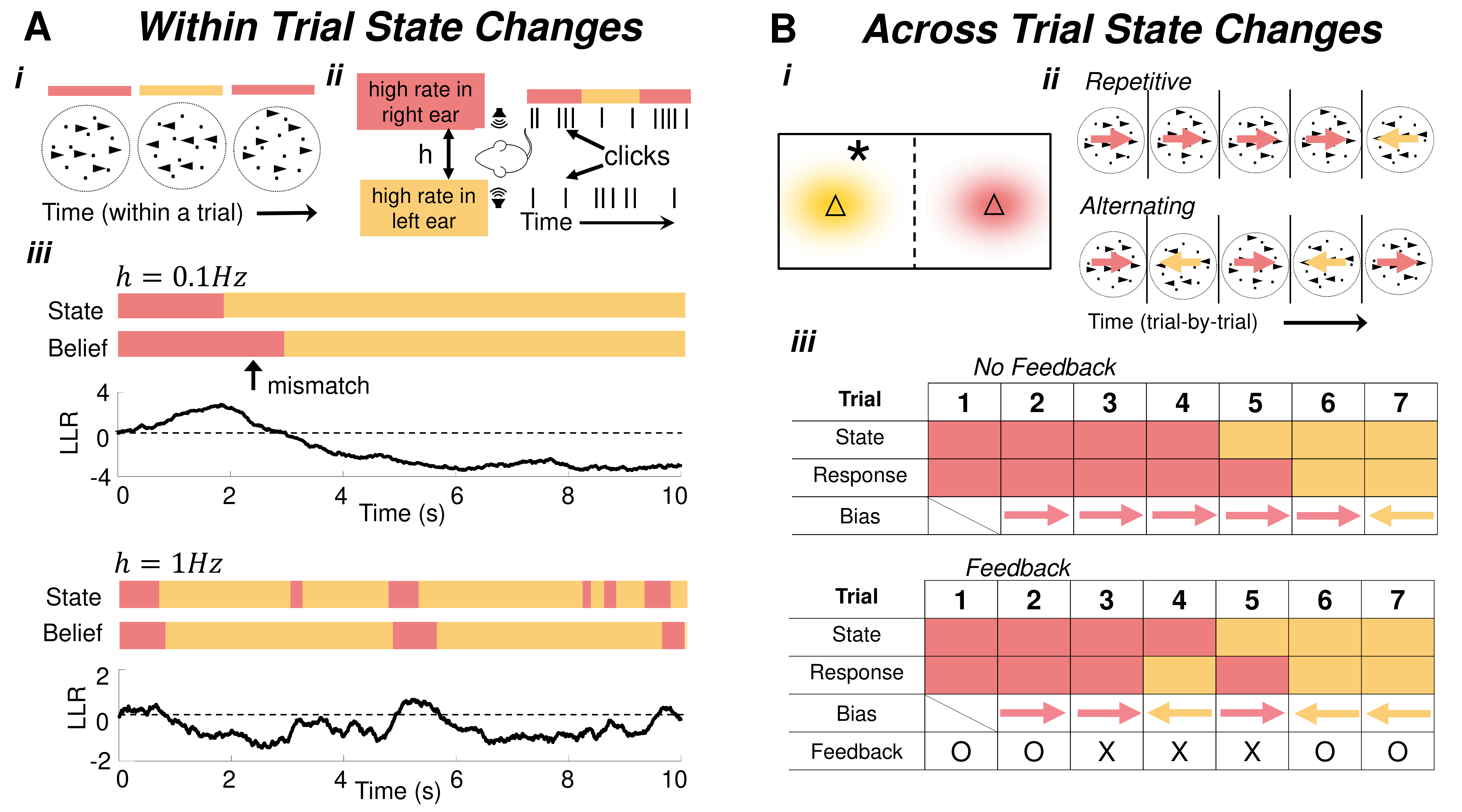} \end{center}
\vspace{-3mm}
\caption{{\bf Dynamic State Changes.}  {\bf (A)} State changes within trials in a {\bf (Ai)} random dot motion discrimination (RDMD) task, in which drift direction switches throughout the trial~\cite{glaze15}, and {\bf (Aii)} dynamic auditory clicks task, in which the side of the higher rate stream alternates during the trial~\cite{piet18}.  {\bf (Aiii)} An ideal observer's LLR (See Eq.~(\ref{SDE}) in \textcolor{blue}{Box 1}) when the hazard rate is low (top panels: h=0.1Hz) and high (bottom panels: h=1Hz). Immediately after state changes,  the belief typically does not match the state. {\bf (B)} State changes across trials. {\bf (Bi)} In the triangles task~\cite{glaze15}, samples (star) are drawn from one of two Gaussian distributions (yellow and red clouds)  whose centers are represented by triangles.  The observers must choose the current center (triangle). {\bf (Bii)} In an RDMD task, dots on each trial move in one of two directions (colored arrows) chosen according to a two-state Markov process. Depending on the switching rate, trial sequences may include excessive repetitions (Top), or alternations (Bottom). {\bf (Biii)}  (Top) Responses can be biased by decisions from previous trials. (Bottom) Probabilistic feedback (`O': correct; `X': incorrect) affects initial bias (e.g., trials 3, 4, and 5), even when not completely reliable. }
\label{fig2:across}
\end{figure}

Normative models of dynamic 2AFC tasks (Fig.~\ref{fig1:schematic}b,c and~\ref{fig2:across}a, Box \textcolor{blue}{1}) exhibit
adaptive, nonlinear discounting of prior beliefs at a rate adapted to expectations of the environment's volatility (Fig.~\ref{fig1:schematic}c), and saturation of certainty about each hypothesis, regardless of how much evidence is accumulated (Fig.~\ref{fig2:across}a). In contrast, ideal observers in static environments weigh all past observations equally, and their certainty grows without bound until a decision~\cite{bogacz06,gold07}.
Also, in dynamic environments, the performance of ideal observers at change points -- times when the correct choice switches -- depends sensitively on environmental volatility (Fig.~\ref{fig2:across}aiii). In slowly changing environments, optimal observers assume that  changes are rare, and thus adapt slowly after one has occured. In contrast, in rapidly changing environments, observers quickly  update their belief after a change point.

The responses of humans and other animals on tasks in which the correct choice changes stochastically during a trial share features with normative models: In a random dot-motion discrimination (RDMD) task, where the motion direction switches at unsignaled changepoints, humans adapt their decision-making process to the switching (hazard) rate (Fig.~\ref{fig2:across}ai)~\cite{glaze15}. However, on average, they overestimate the change rates of rapidly switching environments and underestimate the change rates of slowly switching environments. In a related experiment (Fig~\ref{fig2:across}aii), rats were trained to identify which of two Poisson auditory click streams arrived at a higher rate~\cite{brunton13}. When the identity of the higher-frequency stream switched unpredictably during a trial, trained rats discounted past clicks near-optimally on average, suggesting they learned to account for latent environmental dynamics~\cite{piet18}.

However, behavioral data are not uniquely explained by normative models. Linear approximations of normative models perform nearly identically~\cite{velizcuba16}, and, under certain conditions, fit behavioral data well~\cite{Ossmy13,glaze15,piet18}. Do subjects implement normative decision policies or simpler strategies that approximate them?  Subjects' decision strategies can depend strongly on task design and vary across individuals~\cite{glaze15,glaze18}, suggesting a need for sophisticated model selection techniques. Recent research suggests normative models can be robustly distinguished from coarser approximations when task difficulty and volatility are carefully tuned~\cite{tavoni18}.

\section*{Subjects account for correlations between trials by biasing initial beliefs}
Natural environments can change over  timescales that encompass multiple decisions. However, in many experimental studies, task parameters are fixed or generated independently across trials, so evidence from previous trials is irrelevant. Even so, subjects often use decisions and information from earlier trials to (serially) bias future choices~\cite{fernberger20,frund14,urai17}, reflecting ingrained assumptions about cross-trial  dependencies~\cite{Yu08,nguyen19}.

To understand how subjects adapt to constancy and flux across trials, classic 2AFC experiments have been extended to include correlated cross-trial choices (Fig.~\ref{fig2:across}b) where both evidence accumulated during a trial, and probabilistic reward provide information that can be used to guide subsequent decisions~\cite{kim17,hermoso18}. When a Markov process~\cite{Anderson60} (Fig.~\ref{fig1:schematic}b) is used to generate correct choices, human observers adapt to these trial-to-trial correlations and their response times are accurately modeled by drift diffusion~\cite{goldfarb12} or ballistic models~\cite{kim17} with biased initial conditions. 
 
Feedback or decisions across correlated trials impact different aspects of normative models~\cite{white14} including accumulation speed (drift)~\cite{ratcliff85,diederich06,urai18}, decision bounds~\cite{goldfarb12}, or the initial belief on subsequent trials~\cite{Olianezhad16,purcell16,Braun18}. Given a sequence of dependent but statistically identical trials, optimal observers should adjust their initial belief and decision threshold~\cite{kim17,nguyen19}, but not their accumulation speed in cases where difficulty is fixed across trials~\cite{drugowitsch12}.  Thus, optimal models predict that observers should, on average, respond more quickly, but not more accurately~\cite{nguyen19}. Empirically, humans~\cite{purcell16,Olianezhad16,Braun18} and other animals~\cite{hermoso18} do indeed often respond faster on repeat trials, which can often be modeled by per trial adjustments in initial belief. 
Furthermore, this bias can result from explicit feedback or subjective estimates, as demonstrated in studies where no feedback is provided (Fig.~\ref{fig2:across}biii)~\cite{kim17,Braun18}.

The mechanism by which human subjects carry information across trials remains unclear. Different models fit to human subject data have represented intertrial dependencies using initial bias, changes in drift rate, and updated decision thresholds~\cite{goldfarb12,kim17,urai18}. Humans also tend to have strong preexisting repetition biases, even when such biases are suboptimal~\cite{fernberger20,frund14,urai17}. Can this inherent bias be overcome through training? The answer may be attainable by extending the training periods of humans or nonhuman primates~\cite{glaze15,glaze18}, or using novel auditory decision tasks developed for rodents~\cite{piet18,hermoso18}. Ultimately, high throughput experiments may be needed to probe how ecologically adaptive evidence accumulation strategies change with training.

\section*{Time-varying thresholds account for heterogeneities in task difficulty}

\begin{figure}[t]
\begin{center} \includegraphics[width=9.5cm]{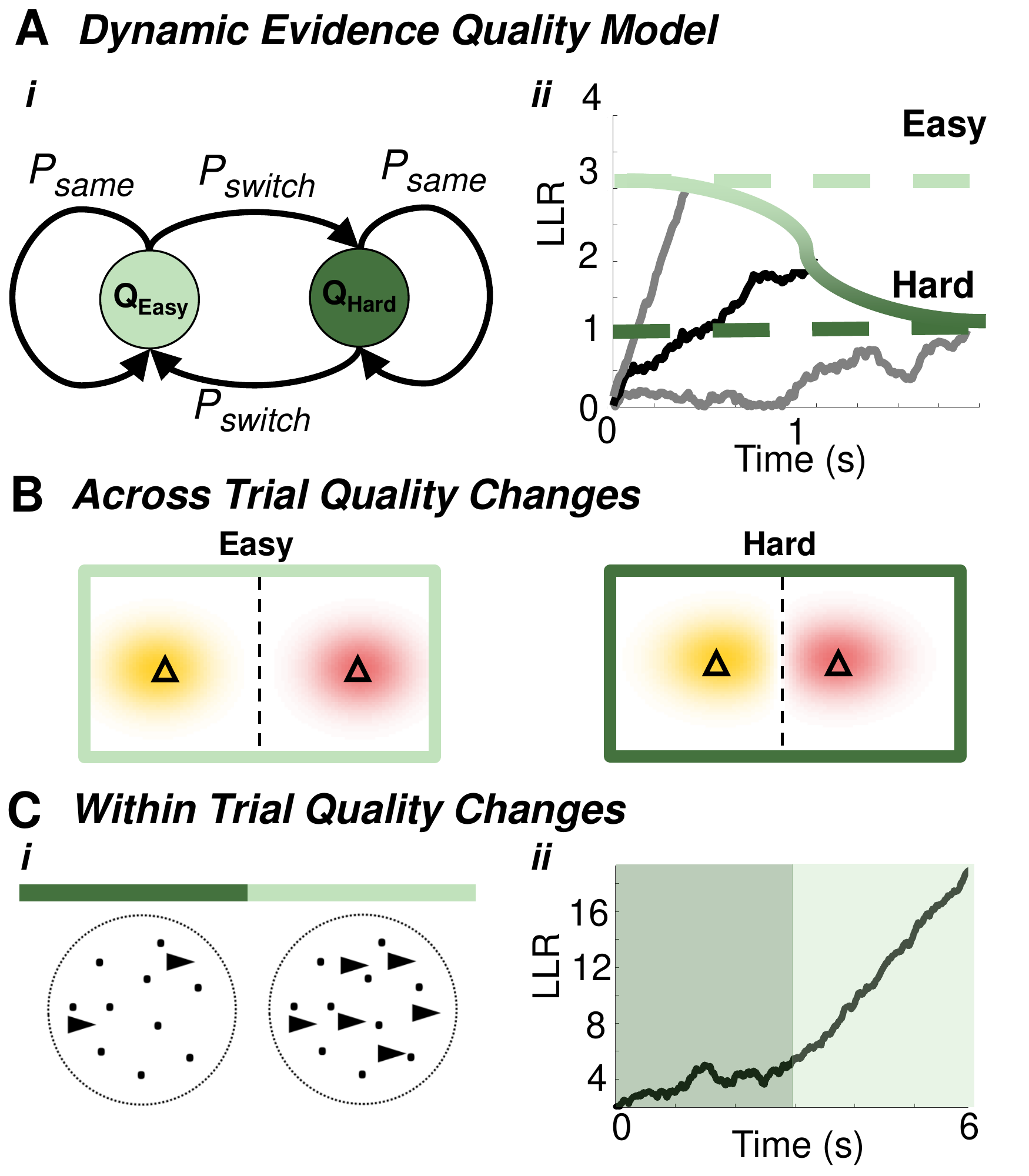} \end{center}
\vspace{-3mm}
\caption{{\bf Dynamic Evidence Quality.} {\bf (A)} Trial-to-trial two-state Markovian evidence quality switching: {\bf (Ai)}  Evidence quality switches between easy ($Q_{\rm easy}$) and hard ($Q_{\rm hard}$) with probability $P_{\text{switch}}$. {\bf (Aii)} Optimal decision policies require time-varying decision thresholds. An observer who knows the evidence quality (easy or hard) uses a fixed threshold (grey traces, dashed lines) to maximize reward rate, but thresholds must vary when evidence quality is initially unknown (black trace, green gradient). {\bf (B)} Different triangle task difficulties (from Fig.~\ref{fig2:across}Ai): Triangles are spaced further apart in easy trials compared to hard trials. {\bf (C)} Changes in quality within trials: {\bf (Ci)} An RDMD task in which the drift coherence increases mid-trial, providing stronger evidence later in the trial. {\bf (Cii)} The corresponding LLR increases slowly early in the trial, and more rapidly once evidence becomes stronger. }
\label{fig3}
\end{figure}

Optimal decision policies can also be shaped by unpredictable changes in decision difficulty. For instance, task difficulty can be titrated by varying the signal-to-noise ratio of the stimulus, so more observations are required to obtain the same level of certainty. Theoretical studies have shown that it is optimal to change one's decision criterion \emph{within} a trial when the difficulty of a decision varies \emph{across} trials~\cite{drugowitsch12,deneve12,malhotra17}. The  threshold that determines how much evidence is needed to make a decision should vary during the trial (Fig.~\ref{fig3}a) to incorporate up-to-date estimates of trial difficulty~\cite{drugowitsch12}. There is evidence that subjects use time-varying decision boundaries to balance speed and accuracy on such tasks~\cite{zhang14,purcell16b}.

Dynamic programming can be used to derive optimal decision policies when trial-to-trial difficulties or reward sizes change. For instance, when task difficulty changes across trials in a RDMD task, optimal decisions are modeled by a DDM with a time-varying boundary, in agreement with reaction time distributions of humans and monkeys~\cite{drugowitsch12,zhang14}. Both dynamic programming~\cite{drugowitsch12} and parameterized function~\cite{thura12,zhang14} based models suggest decreasing bounds maximize reward rates (Fig.~\ref{fig3}a,b). This dynamic criterion helps participants avoid noise-triggered early decisions or extended deliberations~\cite{drugowitsch12}. An exception to this trend was identified in trial sequences without trials of extreme difficulty~\cite{malhotra17}, in which case the optimal strategy used a threshold that increased over time.

Time-varying decision criteria also arise when subjects perform tasks where information quality changes within trials (Fig.~\ref{fig3}c)~\cite{thura12}, especially when  initially weak evidence is followed by stronger evidence later in the trial. However, most studies use heuristic models to explain psychophysical data~\cite{holmes2016new,holmes2018bayesian}, suggesting a need for normative model development in these contexts. Decision threshold switches have also been observed in humans performing changepoint detection tasks, whose difficulty changes from trial-to-trial~\cite{johnson17}, and in a model of value-based decisions, where the reward amounts change between  trials~\cite{tajima16}. Overall,  optimal performance on tasks in which reward structure or decision difficulty changes across trials require time-varying decision criteria, and subject behavior approximates these normative assumptions.

One caveat is that extensive training or obvious across-trial changes are needed for subjects to learn optimal solutions. A meta-analysis of multiple studies showed that fixed threshold DDMs fit human behavior well when difficulty changes between trials were hard to perceive~\cite{hawkins15}. A similar conclusion holds when changes occur within trials~\cite{evans17}. However, when nonhuman primates are trained extensively on tasks where difficulty variations were likely difficult to perceive, they appear to learn a time-varying criterion strategy~\cite{hawkins15b}.  Humans also exhibit time-varying criteria in reward-free trial sequences where interrogations are interspersed with free responses~\cite{palestro18}. Thus, when task design makes it difficult to perceive task heterogeneity or learn the optimal strategy, subjects seem to use fixed threshold criteria~\cite{hawkins15,evans17}. In contrast, with sufficient training~\cite{hawkins15b}, or when changes are easy to perceive~\cite{palestro18}, subjects can learn adaptive threshold strategies.

Questions remain about how well normative models describe subject performance when difficulty changes across or within trials. How distinct do task difficulty extremes need to be for subjects to use optimal models? No systematic study has quantified performance advantages of time-varying decision thresholds. If they do not confer a significant advantage, the added complexity of dynamic thresholds may discourage their use.

\section*{When and how are normative computations learned and achieved?}

Except in simple situations, or with overtrained animals,  subjects can at best approximate computations of an ideal observer~\cite{Geisler03}.  Yet, the studies we reviewed suggest that subjects often learn to do so effectively.
Humans appear to use a process resembling reinforcement learning to learn the structure and parameters of decision task environments~\cite{khodadadi2017learning}. Such learning tracks a gradient in reward space, and subjects adapt rapidly when the task structure changes~\cite{drugowitsch2015tuning}. Subjects also switch between different near-optimal models when making inferences, which may reflect continuous task structure learning~\cite{glaze18}.  However, these learning strategies appear to rely on reward and could be noisier when feedback is probabilistic or absent. Alternatively, subjects may ignore feedback and learn from evidence accumulated within or across trials~\cite{nguyen19,palestro18}.

Strategy learning can be facilitated by using simplified models. For example, humans appear to use sampling strategies that approximate, but are simpler than, optimal inference~\cite{Wilson10,glaze18}.
Humans also behave in ways that limit performance by, for instance, not changing their mind when faced with new evidence~\cite{bronfman15}. This confirmation bias may reflect interactions between decision and attention related systems that are difficult to train away~\cite{talluri18}. Cognitive biases may also arise due to suboptimal applications of normative models~\cite{Beck12}. For instance, recency bias can reflect an incorrect assumption of trial dependencies~\cite{Feldman66}.  Subjects seem to continuously update latent parameters (e.g., hazard rate), perhaps assuming that these parameters are always changing~\cite{Yu08,hermoso18}. 

The adaptive processes we have discussed occur on disparate timescales, and thus likely involve neural mechanisms that interact across scales. Task structure learning occurs over many sessions (days), while the volatility of the environment and other latent parameters can be learned over many trials (hours)~\cite{Wilson10,piet18}. Trial-to-trial dependencies likely require memory processes that span minutes, while within trial changes require much faster adaptation (milliseconds to seconds).

This leaves us with a number of questions: How does the brain bridge timescales to learn and implement adaptive evidence integration? This likely requires coordinating fast neural activity changes with slower changes in network architecture~\cite{radillo17}. Studies of decision tasks in static environments suggest that a subject's belief and ultimate choice is reflected in evolving neural activity~\cite{britten92,bogacz06,gold07,hanks14}. It is unclear whether similar processes represent adaptive evidence accumulation, and, if so, how they are modulated.

\section*{Conclusions}

As the range of possible descriptive models grows with task complexity~\cite{Wilson10,radillo17}, optimal observer models provide  a framework for interpreting behavioral data~\cite{glaze15,piet18,urai18}.  However, understanding the computations subjects use on dynamic tasks, and when they depart from optimality, requires both careful comparison of models to data and comparisons between model classes~\cite{Wu18}.

While we mainly considered optimality defined by performance, model complexity may be just as important in determining the computations used by experimental subjects~\cite{bialek01}. Complex models, while more accurate,  may be difficult to learn, hard to implement, and offer little advantage over simpler ones~\cite{glaze18,radillo17}. Moreover, predictions of more complex models typically have higher variance, compared to the higher bias of more parsimonious models, resulting in a trade-off between the two~\cite{glaze18}.  

Invasive approaches for probing adaptive evidence accumulation are a work in progress~\cite{thura17,akrami18}. However, pupillometry has been shown to reflect arousal changes linked to a mismatch between expectations and observations in dynamic environments~\cite{nassar12,urai17,krishnamurthy17}. Large pupil sizes reflect high arousal after a perceived change, resulting in adaptive changes in evidence weighting.  Thus, pupillometry may provide additional information for identifying computations underlying adaptive evidence accumulation.

Understanding how animals make decisions in volatile environments requires careful task design. Learning and implementing an adaptive evidence accumulation strategy needs to be both rewarding and sufficiently simple so subjects do not resign themselves to simpler computations~\cite{hawkins15,evans17}.  A range of studies have now shown that mammals can learn to use adaptive decision-making strategies in dynamic 2AFC tasks~\cite{glaze15,piet18}. Building on these approaches, and using them to guide invasive studies with mammals offers promising new ways of understanding the neural computations that underlie our everyday decisions.

\section*{Acknowledgements}

We are grateful to Joshua Gold, Alex Piet, and Nicholas Barendregt for helpful feedback.
This work was supported by an NSF/NIH CRCNS grant (R01MH115557) and an NSF grant (DMS-1517629). ZPK was also supported by an NSF grant (DMS-1615737). KJ was also supported by NSF grant DBI-1707400. WRH was supported by NSF grant SES-1556325.

\bibliographystyle{naturemag-doi} 
\bibliography{optimal}
\end{document}